\documentclass[11pt,a4paper]{article}
\usepackage[utf8]{inputenc}
\usepackage{amsmath}
\usepackage{amsfonts}
\usepackage{hyperref}
\usepackage{amssymb}
\usepackage{makeidx}
\usepackage{cite}
\usepackage{bm}
\usepackage{geometry}
\usepackage{braket}
\usepackage[margin=20pt,small]{caption}
\usepackage{tkz-euclide}
\usetkzobj{all}
\usepackage{tikz}
\usetikzlibrary{shapes,arrows,shadows}
\usepackage{doc}

\usepackage{url}

\usepackage{graphicx}
\usepackage{epstopdf}
\begin{document}
\begin{titlepage}

~~\\

\vspace*{0cm}
    \begin{Large}
    \begin{bf}
       \begin{center}
         {Spectrum of Modular Hamiltonian in the Vacuum and Excited States}
       \end{center}
    \end{bf}   
    \end{Large}
    
  \vspace{0.7cm}
\begin{center}   
\bf{Suchetan Das}\footnote
            {
e-mail address : 
suchetan.das@rkmvu.ac.in},
Bobby Ezhuthachan\footnote
            {
e-mail address : 
bobby.ezhuthachan@rkmvu.ac.in}

\vspace{0.3cm}
 {\it Ramakrishna Mission Vivekananda Educational and Research Institute, Belur Math, Howrah-711202, West Bengal, India}
\end{center}
\begin{abstract}

\noindent We study the non-zero eigenmodes for the modular Hamiltonian in the context of AdS$_3/$CFT$_2$. We show how to perturbatively construct zero eigenmodes for the modular Hamiltonian of the class of excited states constructed in \cite{Lashkari:2018oke}, using information about the vacuum non-zero eigen modular modes.

\end{abstract}
\newpage

 \end{titlepage}

\tableofcontents

\rule{\textwidth}{.5pt}\\

\pagenumbering{arabic}

\section{Introduction}\label{I}

The study of the Modular operator ($\Delta$) \cite{Witten:2018lha}-\cite{Borchers:2000pv} has received much attention recently in QFT and also in the context of holography. In the algebraic formulation of QFT, It is a self adjoint positive operator, which generates the automorphism group of the Von Neumann algebra of observables associated with a region $\mathcal{D}(\sigma)$, where $\sigma$ is a spatial region and $\mathcal{D}(\sigma)$ denotes the causal domain of dependence of $\sigma$. Alternatively, in QFT defined with a cut-off, where it is assumed that the Hilbert space can be factorized into $\mathcal{H} =\mathcal{H}_{\sigma}\otimes \mathcal{H}_{\sigma'}$, it is equivalently defined in terms of the reduced density matrix ($\rho$) of the region ($\sigma$) and its compliment ($\sigma'$) as  ($\Delta = \rho_{\sigma}\otimes\rho^{-1}_{\sigma'}$)
\footnote {More precisely, $ln\Delta = ln(\rho_{\sigma}) -ln(\rho_{\sigma'})$.}.
 In QFT the modular operator\footnote{More precisely, a variant- the relative modular operator} or its natural logarithm ($K = -ln(\Delta)$)\footnote{$K$ can be expressed as $K_{\sigma}= H_{\sigma} -H_{\sigma'}$, where $H_{\sigma} = -ln(\rho_{\sigma})$, in the case when the Hilbert space is assumed to be factorized.} have been used to give rigorous entropy bounds \cite{Casini:2008cr}-\cite{Bousso:2014uxa}, to prove generalized second laws \cite{Wall:2011hj} and to constrain correlators in QFT \cite{Lashkari:2018nsl} which have, among other interesting results, lead to proofs for various null energy conditions \cite{Faulkner:2016mzt}-\cite{Ceyhan:2018zfg}.

Its usefulness in holography comes from its identification with its bulk counterpart \cite{Jafferis:2014lza}. An identification which is expressed mathematically as an equality of the bulk and boundary `Total Modular Hamiltonian' ($K$) at large $N$ \cite{Jafferis:2015del},\cite{Harlow:2016vwg}. 
\begin{equation}\label{JLMS}
K_{\sigma} = K_{\Sigma} +\mathcal{O}(1/N)
\end{equation}
Here $\Sigma$ corresponds to the spatial region bounded by $\sigma$ and the Ryu-Takayanagi(RT) surface\cite{Ryu:2006bv},\cite{Hubeny:2007xt}. This identification can be used to express operators inside $\mathcal{D}(\Sigma)$ in terms of the modular transformed operators ($\mathcal{O}_s = \Delta^{-is}\mathcal{O}\Delta^{is}$) in the boundary region ($\sigma$) \cite{Faulkner:2017vdd}and thus provides evidence for `Entanglement Wedge Reconstruction' \cite{Jafferis:2015del},\cite{Czech:2012bh}-\cite{Harlow:2018fse} in holography. Equation(\ref{JLMS}), has been used in \cite{Kabat:2017mun}-\cite{Kabat:2018smf} to give an alternate derivation of the HKLL formula \cite{Hamilton:2005ju}-\cite{Hamilton:2006fh} for the bulk scalar field in $AdS_3$ and the BTZ blackhole background\footnote{This was then generalized for an infinite class of geometries, which are locally AdS$_3$ and related to AdS$_3$ by a large diffeomorphism \cite{Das:2018ojl}}. In this context, zero-eigenmodes of the Modular Hamiltonian has played an important role. Eigenmodes of the Modular Hamiltonian are defined as follows \cite{Faulkner:2017vdd}.
\begin{equation}
[K, \mathcal{O}_{k}] = k\mathcal{O}_{k}
\end{equation}

In particular, the zero eigenmodes are given by $k=0$. Such eigenmodes can be constructed from operators ($\mathcal{O}$) with support inside $\mathcal{D}(\sigma)$, by taking the Fourier transforms of the modular transform of such operators.
\begin{equation}\label{mode defn}
\mathcal{O}_{k} = \int ds \; e^{-iks} \; \mathcal{O}_s
\end{equation}
In \cite{Faulkner:2017vdd}, it was argued that the scalar zero mode can be expressed as an integral over the RT surface of the bulk scalar field. 
\begin{equation}\label{zero-mode bulk}
\mathcal{O}_{0}(x) = 4\pi \int_{\text{RT}}dY_{\text{RT}}\braket {\phi(Y_{\text{RT}})\mathcal{O}(x)} \phi(Y_{\text{RT}})
\end{equation}  

Interestingly, for the case of the vacuum state in $CFT_2$, with $\sigma$ corresponding to a single spatial interval, the (scalar) OPE blocks provide an important example of the zero eigen modes \cite{Czech:2016xec}-\cite{Czech:2017zfq}. Along with equation (\ref{zero-mode bulk}), this fact then provides a bulk formula for these OPE blocks in terms of geodesic integral of bulk scalar fields.

In this note, we take a closer look at the eigenmodes in the $CFT_2$ example. In particular, we look at the non-zero modes in this case. In section \ref{II}, we observe, that OPE blocks, constructed out of an OPE of two operators with non vanishing spin differences are non zero eigen modes, with the difference in spins being the eigenvalue for these non zero mode. This observation then leads us to a dual picture for these modes in the bulk. The details of this  bulk construction is spelt out in section \ref{III}. This generalizes the discussion on OPE blocks as zero modes and their duals description as geodesic operators. In section \ref{IV}, we use the results of \cite{Lashkari:2018oke}, to find the zero modes in excited states as a perturbation expansion around the vacuum eigenmodes. We show how to systematically extract the higher order corrections, using the knowledge of the full spectrum of vacuum eigenmodes. Details of calculations and related reviews of previous works have been added in appendices.

\section{Modular eigenmodes in Vacuum CFT$_{2}$}\label{II}
 
In  CFT$_2$, the scalar OPE blocks form a useful basis for the zero modes as discussed in \cite{Czech:2017zfq}. In this section we point out that a different class of the OPE blocks also provide examples of non-zero eigenmodes. 
   
In spherical subregions for the vacuum state in a CFT, the modular Hamiltonian has a local expression in terms of stress tensor\cite{Hislop:1981uh},\cite{Casini:2011kv}. In particular, the modular Hamiltonian of a single interval $R$:[($y_{1},\bar{y}_{1}$),($y_{2},\bar{y}_{2}$)] in a 2D CFT is given by, 
\begin{align}\label{modvac}
H_{R} = \int_{y_{1}}^{y_{2}} d\omega \frac{(\omega - y_{1})(y_{2}-\omega)}{y_{2}-y_{1}} T_{\omega\omega}(\omega) + \int_{\bar{y}_{1}}^{\bar{y}_{2}} d\bar{\omega} \frac{(\bar{\omega} - \bar{y}_{1})(\bar{y}_{2}-\bar{\omega})}{\bar{y}_{2}-\bar{y}_{1}} \bar{T}_{\bar{\omega}\bar{\omega}}(\bar{\omega})
\end{align}
Then $K$ is given by $H_R$ - $H_{R'}$, with $R'$ being the complimentary region.  
 
We first study the action of $K$\footnote{From now, instead of calling $K$ as `full modular Hamiltonian', we will simply denote it as modular Hamiltonian.} on all possible OPE blocks of dimension $h_{k} \neq \bar{h}_{k}$, constructed out of OPE of two operators with unequal non-vanishing spin i.e ($h_{i},\bar{h}_{i}) \neq (h_{j},\bar{h}_{j}$) as well as $h_{i}\neq \bar{h}_{i}$,$h_{j}\neq \bar{h}_{j}$. In CFT$_{2}$, using Shadow operator formalism \cite{Ferrara:1973vz}-\cite{SimmonsDuffin:2012uy}, it is easy to find an integral expression for such OPE blocks\cite{Czech:2016xec},\cite{Das:2018ajg}.
\begin{align}
B^{ij}_{k}(y_{1},\bar{y}_{1};y_{2},\bar{y}_{2}) =& n_{ijk}\int_{y_{1}}^{y_{2}}d\zeta \int_{\bar{z}_{1}}^{\bar{z}_{2}}d\bar{\zeta}\left(\frac{(\zeta-y_{1})(y_{2}-\zeta)}{y_{2}-y_{1}}\right)^{h_{k}-1}\left(\frac{y_{2}-\zeta}{\zeta-y_{1}}\right)^{h_{ij}}\times \nonumber \\
&\left(\frac{(\bar{\zeta}-\bar{y}_{1})(\bar{y}_{2}-\bar{\zeta})}{\bar{y}_{2}-\bar{y}_{1}}\right)^{\bar{h}_{k}-1}\left(\frac{\bar{y}_{2}-\bar{\zeta}}{\bar{\zeta}-\bar{y}_{1}}\right)^{\bar{h}_{ij}}\mathcal{O}_{k}(\zeta,\bar{\zeta})
\end{align}
Using $h_{i}=\frac{\Delta_{i}+l_{i}}{2}$ and $\bar{h}_{i}=\frac{\Delta_{i}-l_{i}}{2}$, the expression becomes,
\begin{align}\label{opeblocks}
B^{ij}_{k}(y_{1},\bar{y}_{1};y_{2},\bar{y}_{2}) =& n_{ijk}\int_{y_{1}}^{y_{2}}d\zeta \int_{\bar{z}_{1}}^{\bar{z}_{2}}d\bar{\zeta}\left(\frac{(\zeta-y_{1})(y_{2}-\zeta)}{y_{2}-y_{1}}\right)^{h_{k}-1}\left(\frac{(\bar{\zeta}-\bar{y}_{1})(\bar{y}_{2}-\bar{\zeta})}{\bar{y}_{2}-\bar{y}_{1}}\right)^{\bar{h}_{k}-1}\times \nonumber \\
&\left(\frac{(y_{2}-\zeta)(\bar{y}_{2}-\bar{\zeta})}{(\zeta-y_{1})(\bar{\zeta}-\bar{y}_{1})}\right)^{\frac{\Delta_{ij}}{2}}\left(\frac{(y_{2}-\zeta)(\bar{\zeta}-\bar{y}_{1})}{(\zeta-y_{1})(\bar{y}_{2}-\bar{\zeta})}\right)^{\frac{l_{ij}}{2}}\mathcal{O}_{k}(\zeta,\bar{\zeta})
\end{align}
To compute $[K,B^{ij}_{k}]$, we first need to determine the action of $K$ on primary $\mathcal{O}_{k}$. To do this, we use $T\mathcal{O}_{k}$ OPE to get the following,
\begin{align}
2\pi[T(\omega),\mathcal{O}_{k}(\zeta,\bar{\zeta})] = 2\pi i(h\partial_{\zeta}\delta(\zeta-\omega)+\delta(\zeta-\omega)\partial_{\zeta})\mathcal{O}_{k}(\zeta,\bar{\zeta}), \nonumber \\
2\pi[\bar{T}(\bar{\omega}),\mathcal{O}_{k}(\zeta,\bar{\zeta})] = -2\pi i(h\partial_{\bar{\zeta}}\delta(\bar{\zeta}-\bar{\omega})+\delta(\bar{\zeta}-\bar{\omega}) \partial_{\bar{\zeta}})\mathcal{O}_{k}(\zeta,\bar{\zeta})
\end{align}
Since $K$ is the sum of right moving($K^{(R)}(y_{1},y_{2})$) and left moving($K^{(L)}(\bar{y}_{1},\bar{y}_{2})$) part, involving $T$ and $\bar{T}$ respectively, one can easily compute the following commutators using previous $[T,\mathcal{O}]$ commutator.
\begin{align}\label{ko commutator}
[K^{(R)},\mathcal{O}_{k}(\zeta,\bar{\zeta})] &= \frac{2\pi i}{y_{2}-y_{1}} \left( h_{k}(y_{2}+y_{1}-2\zeta)+ (\zeta-y_{1})(y_{2}-\zeta)\partial_{\zeta}\right)\mathcal{O}(\zeta,\bar{\zeta}), \nonumber \\
[K^{(L)},\mathcal{O}(\zeta,\bar{\zeta})] &= -\frac{2\pi i}{\bar{y}_{2}-\bar{y}_{1}} \left( \bar{h}_{k}(\bar{y}_{2}+\bar{y}_{1}-2\bar{\zeta})+ (\bar{\zeta}-\bar{y}_{1})(\bar{y}_{2}-\bar{\zeta})\partial_{\bar{\zeta}}\right)\mathcal{O}(\zeta,\bar{\zeta})
\end{align}
Using these relations, one could easily compute the following commutators upto vanishing total derivatives\footnote{The details are given in appendix (\ref{VI}).}
\begin{align}\label{main commutator}
[K^{(R)},B^{ij}_{k}] = \pi i(\Delta_{ij}+l_{ij})B^{ij}_{k}; \quad [K^{(L)},B^{ij}_{k}] = \pi i(l_{ij}-\Delta_{ij})B^{ij}_{k}
\end{align}
Hence we get our desired commutator
\begin{align}
[K,B^{ij}_{k}] = 2\pi il_{ij}B_{k}^{ij}
\end{align}
Hence, the non zero modes of $K$ are precisely the OPE blocks constructed out of OPE with non zero spin difference $l_{ij}$. Another interesting quantity is the antisymmetric combination of those left and right moving components of $K$, which is called $P_{D}=K^{(R)}-K^{(L)}$ \cite{Czech:2017zfq}. Hence, the commutator with $P_{D}$ gives
\begin{align}
[P_{D},B^{ij}_{k}] = 2\pi i\Delta_{ij}B_{k}^{ij}
\end{align}
The importance of $P_{D}$ will be discussed in the next section where we are going to construct the bulk dual of non-zero modes.

\section{AdS$_{3}$ dual of non zero modular modes in vacuum}\label{III}
In CFT$_{2}$, the scalar OPE blocks(as well as spinning blocks) constructed out of an OPE of two scalar operators(or spinning operators with equal spin) are zero modes of the Modular Hamiltonian. The dual description is already known in the literature in terms of geodesic operators in AdS$_{3}$. Scalar geodesic operators $B_{k}$ are defined as bulk scalar field $\phi(x,z,t)$ integrated over the geodesic $\lambda$ (with infinitesimal length $ds$) anchored on the boundary points $x_{1}$,$x_{2}$, where the two operators of the OPE.
\begin{align}
B_{k}(x_{1},x_{2}) = \int_{\lambda} ds \phi(x(s),z(s),t(s))
\end{align}
If those two operators have unequal dimensions($\Delta_{i} \neq \Delta_{j}$), there is an extra exponential weight $e^{-s\Delta_{ij}}$\footnote{$\Delta_{ij} = \Delta_{i}-\Delta_{j}$} contribution to the integral.
\begin{align}\label{modified}
B_{k}^{ij}(x_{1},x_{2}) = \int_{\lambda} ds e^{-s\Delta_{ij}} \phi(x(s),z(s),t(s))
\end{align}
This expression was derived in \cite{daCunha:2016crm}, by rewriting an old formula of OPE blocks in terms of bulk variables. The presence of the weight factor inside the integral, can be understood also by studying modular zero modes under modular flows in boundary and bulk as argued in \cite{Czech:2017zfq}. 
In this section, we propose that the dual for the scalar non-zero modes that we wrote down in the previous section is given by:
\begin{align}\label{generalized expression}
B^{ij}_{k} = c_{k}\int_{\text{cylinder}}d\tilde{t} ds e^{-\tilde{t}l_{ij}}e^{-s\Delta_{ij}} \phi(x(s, \tilde{t}), z(s, \tilde{t}), t(s,\tilde{t}))
\end{align}

Where the integral is now over the Lorentzian cylindrical surface generated by the $P_{D}$ and the $K$. As will be shown in detail below, the $P_{D}$ generates flows along the geodesic direction while the $K$ generates boosts around the geodesic, which vanishes on the geodesic. Thus, when $l_{ij}=0$, we should get back the previous expression which involves an integral along the geodesic. We will also motivate the additional factor of $e^{-l_{ij}\tilde{t}}$ from the action of modular Hamiltonian on modular modes. To obtain the proposed form of non zero modes, we will begin by reviewing the discussion given in \cite{Czech:2017zfq}.


In CFT$_{2}$, let us consider two points at $t=0$ and $x=-R,R$. The associated causal diamond has upper and lower tips $(y^{\mu},x^{\mu})$, where $(y^{0}=R,y^{1}=0),(x^{0}=-R.x^{1}=0)$. The conformal killing vector $\kappa^{\mu}$ which preserves the diamond is given by the following expression,
\begin{align}\label{boundary flow}
\kappa^{\mu}\delta_{\mu} = \frac{\pi}{R}\left[(R^{2}-x^{2}-t^{2})\partial_{t} - 2tx\partial_{x}\right]
\end{align}
One can easily check that the four vertices of the diamond are fixed points of the flow. Also $\kappa^{\mu}$ is timelike and future directed everywhere except at the boundaries of the diamonds where it is null(i.e $\kappa^{2} = 0$). In terms of light cone coordinates $\zeta=x-t, \bar{\zeta}=x+t$, the killing vector has two components $\kappa^{\zeta},\kappa^{\bar{\zeta}}$. Where,
\begin{align}
\kappa^{\zeta} = \pi\frac{(R+\zeta)(\zeta-R)}{R}; \quad \kappa^{\bar{\zeta}} = \pi\frac{(R-\bar{\zeta})(\bar{\zeta}+R)}{R}
\end{align}
Therefore, these two components are generators of $SO(1,1)\times SO(1,1)$ subgroup of the global conformal group $SO(2,2)$, that stabilizes the causal diamond. One can easily recognise that modular Hamiltonian (\ref{modvac}) can be written in a covariant form in terms of $\kappa^{\mu}$
\begin{align}
H_{mod} = \int_{\diamondsuit}d\Sigma^{\mu}\kappa^{\nu}T_{\mu\nu} = \int \kappa^{\zeta}T_{\zeta\zeta}d\zeta + \int \kappa^{\bar{\zeta}}T_{\bar{\zeta}\bar{\zeta}}d\bar{\zeta}
\end{align}
Similarly, $P_{D}$ can be defined as an antisymmetric combination of the generators,
\begin{align}
P_{D} = \int \kappa^{\zeta}T_{\zeta\zeta}d\zeta - \int \kappa^{\bar{\zeta}}T_{\bar{\zeta}\bar{\zeta}}d\bar{\zeta}
\end{align}
In terms of global conformal generators $L_{0}=-\zeta\partial_{\zeta}$,$L_{1}=\zeta^{2}\partial_{\zeta}$,$L_{-1}=\partial_{\zeta}$ and the corresponding left moving generators $\bar{L}_{0}$,$\bar{L}_{1}$ and $\bar{L}_{-1}$ the (anti)symmetric combinations of killing vectors can be written as following:
\begin{align}\label{generators}
\kappa_{\zeta}\partial_{\zeta}+\kappa_{\bar{\zeta}}\partial_{\bar{\zeta}} = (L_{1}-\bar{L}_{1})-R^{2}(L_{-1}-\bar{L}_{-1}) \\
\kappa_{\zeta}\partial_{\zeta}-\kappa_{\bar{\zeta}}\partial_{\bar{\zeta}} = (L_{1}+\bar{L}_{1})-R^{2}(L_{-1}+\bar{L}_{-1})
\end{align}
It is straightforward to see that the action of $H_{mod}$ and $P_{D}$ are perpendicular and using (\ref{generators}) one can check that indeed $[H_{mod},P_{D}]=0$. Hence, $P_{D}$ generates translation along modular time slice, while $H_{mod}$ generates modular flows.

One could extend these CFT$_{2}$ killing vectors to bulk AdS$_{3}$ killing vectors. Since, conformal symmetry in the boundary acts as isometries in the bulk AdS$_{3}$, the bulk generators in Poincare coordinates can be written in the following way:
\begin{align}
L_{b,0}=-\frac{1}{2}z\partial_{z}-\zeta\partial_{\zeta},\quad L_{b,1}=z\zeta\partial_{z}+\zeta^{2}\partial_{\zeta}-z^{2}\partial_{\bar{\zeta}}, \quad L_{b,-1}=\partial_{\zeta}, \\
\bar{L}_{b,0}=-\frac{1}{2}z\partial_{z}-\bar{\zeta}\partial_{\bar{\zeta}},\quad \bar{L}_{b,1}=z\bar{\zeta}\partial_{z}+\bar{\zeta}^{2}\bar{\partial_{\zeta}}-z^{2}\partial_{\zeta}, \quad \bar{L}_{b,-1}=\partial_{\bar{\zeta}}
\end{align}
Therefore the (anti)symmetric combinations of killing vectors are extended to AdS$_{3}$ as follows
\begin{align}\label{H,P}
H_{b,mod}: \quad \kappa_{b,\zeta}\partial_{\zeta}+\kappa_{b,\bar{\zeta}}\partial_{\bar{\zeta}} = (R^{2}-x^{2}-t^{2})\partial_{t}-2xt\partial_{x}-z^{2}\partial_{t}-2zt\partial_{z} \\
P_{b,D}: \quad \kappa_{b,\zeta}\partial_{\zeta}-\kappa_{b,\bar{\zeta}}\partial_{\bar{\zeta}} = -(R^{2}-x^{2}-t^{2})\partial_{x}+2xt\partial_{t}-z^{2}\partial_{x}-2zx\partial_{z}
\end{align}
One can easily check that $P_{b,D}$ generates translation along geodesic in AdS$_{3}$. Since, at $t=0$ and on the geodesic $x^{2}+z^{2}=R^{2}$, $P_{b,D} \sim -2z^{2}\partial_{x}+2xz\partial_{z}$. Hence, the curve generated by $P_{b,D}$ and parametrized by some parameter $s$, satisfies the following two equations,
\begin{align}
\frac{dx(s)}{ds}=-2z^{2}; \quad \frac{dz(s)}{ds}=2xz
\end{align}
Therefore, $xdx+zdz=0$ is the equation of the curve, which is precisely the geodesic equation. Thus $P_{b,D}$ acts as translation along the geodesic. 

The authors of \cite{Czech:2017zfq}\footnote{In\cite{Czech:2017zfq}, the authors study the ``modular Berry transformation"  induced by a change in the length of the single interval. The correpsonding ``modular Berry curvature" has been identified with the bulk Riemann curvature\cite{Czech:2018kvg}, \cite{Czech:2019vih}.} observe that the fact that the zero modes\footnote{which in this case are the OPE blocks made out of OPE of two scalar operators with unequal difference of dimension $\Delta_{ij}$ for which $[H_{mod},B_{k}^{ij}]=0$}, are eigenoperators of $P_{D}$(i.e $[P_{D},B_{k}^{ij}] \propto \Delta_{ij}B^{ij}_{k}$), would imply that they transform as follows under a finite transformation of magnitude $s_{0}$:
\begin{align}\label{translation}
B^{ij}_{k} \rightarrow e^{s_{0}\Delta_{ij}}B^{ij}_{k}
\end{align}
Correspondingly in the bulk, one must incorporate such transformation from the action of $P_{b,D}$ in the bulk. If $s$ is the proper length parameter along the geodesic, translation along $s$ by a shift of $s_{0}$ should result in a transformation similar to (\ref{translation}). This explains the additional weight factor of $e^{-s\Delta_{ij}}$ inside the integral in equation (\ref{modified})

This argument extends to the case of non-zero modes. The fact that $[H_{mod},B^{ij}_{k}] \propto l_{ij}B^{ij}_{k}$, now similarly implies that under finite transformation generated by $H_{mod}$, of magnitude $\tilde{t}_{0}$, the modes transform as $B^{ij}_{k} \rightarrow e^{\tilde{t}_{0}l_{ij}}B^{ij}_{k}$. Since $H_{mod}$ and $P_{D}$ commutes, the full transformation on non zero modes is just product of each transformation,
\begin{align}\label{non zero modes transformation}
B^{ij}_{k} \rightarrow e^{s_{0}\Delta_{ij}}e^{\tilde{t}_{0}l_{ij}}B^{ij}_{k}
\end{align}
This is the reason for the extra weight factors in equation (\ref{generalized expression}). 

We now study explicitly the curves generated by $H_{b,mod}$. 

From (\ref{H,P}), we see the tangents of $t,x,z$ along the curves satisfy the set of equations,
\begin{align}\label{curve equation}
\frac{dt}{ds}=R^{2}-t^{2}-x^{2}-z^{2}; \quad \frac{dx}{ds}=-2xt; \quad \frac{dz}{ds}=-2zt
\end{align}
After solving these equations, we get the final equation of curve generated by $H_{mod}$
\begin{align}\label{z equation}
\eta^{2}t^{2}-(z+R\gamma\eta)^{2} = \eta^{2}R^{2}(1-\gamma^{2})
\end{align}
\begin{align}\label{x equation}
\hat{\eta}^{2}t^{2}-(x+R\gamma\hat{\eta})^{2} = \hat{\eta}^{2}R^{2}(1-\gamma^{2})
\end{align}
For the details of the solution and the definitions of $\eta$,$\hat{\eta}$,$\gamma$, see appendix (\ref{VII}). These equations describe a Lorentzian cylinder around the geodesic. Also we have $\eta^{2}+\hat{\eta}^{2} = 1$. At $t=0$ we have the following two set of equations,
\begin{align}
(z+R\gamma\eta)^{2} = \eta^{2}R^{2}(\gamma^{2}-1) \quad \text{and} \quad (x+R\gamma\hat{\eta})^{2} = \hat{\eta}^{2}R^{2}(\gamma^{2}-1)
\end{align}
In this case, the existence of solution implies that $\gamma^{2} \geq 1$. At $\gamma=1$, $z=-R\eta$ and $x=-R\hat{\eta}$. Hence it satisfies $x^{2}+z^{2}=R^{2}$, the equation for geodesic. Therefore, at $\gamma^{2}=1$, the curve reduces to the constant time geodesic in AdS$_{3}$. But we are interested in the general($t\neq 0$) solution. From (\ref{z equation}) and (\ref{x equation}), using $\eta = -\sin\theta$,$\hat{\eta}= -\cos\theta$ we can parametrize the cylinder in boost parameter $\rho$ and geodesic parameter $\theta$,
\begin{align}\label{defn of z,x,t}
t=R\sqrt{\gamma^{2}-1}\sinh\rho \\
z=R\sin\theta\gamma + R\sin\theta\sqrt{\gamma^{2}-1}\cosh\rho \\
x=R\cos\theta\gamma + R\cos\theta\sqrt{\gamma^{2}-1}\cosh\rho
\end{align}
Therefore, in the bulk, $H_{b,mod}$ generates boost in both $x-t$ and $z-t$ plane. In the boundary, i.e at $\theta=0$, one can easily see the curves are generated by boost in $x-t$ plane.
\begin{align}
(x-R\gamma)^{2}-t^{2} = R^{2}(\gamma^{2}-1)
\end{align}
Solving (\ref{boundary flow}) one could also obtain the above equation which is generated by boundary modular flow. Now it is easy to construct a distance function $\tilde{t}$ from $\rho=\infty$ to some finite value of $\rho$ in the cylinder. It is defined as follows.
\begin{align}
\tilde{t} &= \int^{\rho}_{\infty}\frac{\sqrt{(\frac{dt}{d\rho})^{2}-(\frac{dx}{d\rho})^{2}-(\frac{dz}{d\rho})^{2}}}{z}d\rho = \frac{\sqrt{\gamma^{2}-1}}{\sin\theta}\ln\left[\frac{\sqrt{\gamma^{2}-1}e^{\rho}+\gamma-1}{\sqrt{\gamma^{2}-1}e^{\rho}+\gamma+1}\right]
\end{align}
In contrast to the dual zero mode picture, the non-zero modes transform under boost in the bulk and hence they are not located on the geodesic. From the previous analysis of $H_{b,mod}$, we expect the modes can be located any point on the cylinder. Hence, instead of integrating over the geodesic, we need to integrate over the cylinder to get the non-zero modes. It is expected from the action of stabilizer group $SO(1,1)\times\bar{SO(1,1)}$ in the bulk. The most general object, that is invariant under two $SO(1,1)$s(translation and boost around the geodesic), is the Lorentzian cylinder we described above. Geodesic is only a special curve centered inside such cylinder. This cylinder basically describes a section of entanglement wedge(in the pure AdS, it coincides with causal wedge though). So it is well expected that non zero modes live anywhere in the entanglement wedge, while zero modes live on the RT surface. Also the transformation property of non zero modes (\ref{non zero modes transformation}) suggests an extra exponential weight of $e^{-\tilde{t}l_{ij}}$ contributes in the bulk integral over the cylinder. Hence finally we expect the non zero modes or the OPE blocks made out of operators with non vanishing spin difference are of the form,
\begin{align}\label{main expression}
B^{ij}_{k} = c_{k}\int_{\text{cylinder}}d\tilde{t} ds e^{-\tilde{t}(\rho,\theta)l_{ij}}e^{-s(\theta)\Delta_{ij}} \phi(x(\rho,\theta),t(\rho,\theta),z(\rho,\theta))
\end{align}
Where $c_{k}$ is some normalization constant. We need to show that this form indeed satisfy the following three condition:
 \begin{enumerate}
 \item  Casimir eigenvalue equation on OPE blocks $B^{ij}_{k}$ $\iff$ Klein-Gordon equation of motion on field $\phi$
 \item Boundary condition on $B^{ij}_{k}$ at the coincidence limit $x_{1}\rightarrow x_{2}$ or, $R \rightarrow 0$ $\iff$ AdS/CFT boundary condition on $\phi$ at $z\rightarrow 0$.
 \item When $l_{ij}=0$ , the definition of $B^{ij}_{k}$ reduces to that of geodesic operator as in (\ref{modified}).
 \end{enumerate}
 To show the first condition, we can apply the same argument of intertwinement discussed in \cite{Czech:2016xec},\cite{deBoer:2016pqk}. Let us apply the conformal generator $L_{i}(x_{1})+L_{i}(x_{2})$ to the above expression (\ref{main expression}). The action of the generator shifts the cylinder by infinitesimal displacement along the direction of bulk killing vector field $L_{b,i}$. This displacement can be absorbed in the change of location of the field by an amount $L_{b,i}\phi$ from the value at the original location. But since $L_{b,i}$ is killing vector field, it does not affect the other part of the integrand which is manifestly diffeomorphism invariant. Thus what we get is,
 \begin{align}
 (L_{i}(x_{1})+L_{i}(x_{2}))\int_{\text{cylinder}}ds d\tilde{t} e^{-\tilde{t}l_{ij}}e^{-s\Delta_{ij}} \phi(\rho,\theta) = \int_{\text{cylinder}}ds d\tilde{t} e^{-\tilde{t}l_{ij}}e^{-s\Delta_{ij}} L_{b,i}\phi(\rho,\theta)
 \end{align}
 To get the action of Casimir, we must act $L_{i}(x_{1})+L_{i}(x_{2})$ twice\cite{deBoer:2016pqk}. 
 \begin{align}
 C^{ij} (L_{i}(x_{1})+L_{i}(x_{2})) (L_{j}(x_{1})+L_{j}(x_{2}))\int ds d\tilde{t} e^{-\tilde{t}l_{ij}}e^{-s\Delta_{ij}} \phi = \int ds d\tilde{t} e^{-\tilde{t}l_{ij}}e^{-s\Delta_{ij}} C^{ij}L_{b,i}L_{b,j}\phi
 \end{align}
 One can see immediately that $C^{ij}L_{b,i}L_{b,j}$ is proportional to Laplacian acting in AdS$_{3}$.
 \begin{align}
 C^{ij}L_{b,i}L_{b,j}\phi = -\nabla^{2}\phi = -m^{2}\phi = -\Delta(\Delta-1)\phi
 \end{align}
 Thus the first condition is ensured. To proceed with the next check on boundary condition, we first recall the AdS/CFT boundary condition- $\phi(z\rightarrow 0,x) \sim z^{\Delta}\mathcal{O}_{\Delta}(x)$. At the coincident point, the OPE block reduces to local operator as a leading contribution, i.e $B^{ij}_{k}(x_{1}\rightarrow x_{2}) \sim |x_{1}-x_{2}|^{\Delta_{k}}\mathcal{O}_{k}(x_{1})$. The coincidence limit $R \rightarrow 0$, is the limit in which the bulk field approaches the boundary. Hence using (\ref{defn of z,x,t}), definition of $\tilde{t}$, $s=-\ln(\csc\theta+\cot\theta)$ and applying AdS/CFT boundary condition we get,
 \begin{align}
& \lim_{R \rightarrow 0} B^{ij}_{k}(R,-R) \nonumber \\
& = c_{k}R^{\Delta_{k}}\int d\rho d\theta (\csc\theta+\cot\theta)^{\Delta_{ij}}(\sin\theta)^{\Delta_{k}-2}(\gamma+\sqrt{\gamma^{2}-1}\cosh\rho)^{\Delta_{k}-1} \nonumber \\
&\times \sqrt{\gamma^{2}-1}\left[\frac{\sqrt{\gamma^{2}-1}e^{\rho}+\gamma-1}{\sqrt{\gamma^{2}-1}e^{\rho}+\gamma+1}\right]^{ \frac{\sqrt{\gamma^{2}-1}}{\sin\theta}l_{ij}} \mathcal{O}_{k}(R) \nonumber \\
& = \tilde{c}_{k} R^{\Delta_{k}}\mathcal{O}_{k}(R)
 \end{align}
 Where $\tilde{c}_{k}$ is the new normalization constant comes from $\rho$ and $\theta$ integral. Therefore, from boundary condition we can fix the normalization constant. But from these two conditions, we can not fix the last condition i.e reduction to geodesic operator when $l_{ij} = 0$. One can immediately see that for $l_{ij}=0$,  $\gamma^{2} = 1$ since it is for this value of $\gamma^{2}$, that  the cylinder collapses to the geodesic. But since $\gamma$ is a constant appearing when we solve the equations of curve(see appendix (\ref{VII})), we have freedom to choose it. The simplest choice would be $l_{ij} = (\gamma^{2}-1)$.  
 
 \section{Modular eigenmodes in excited states}\label{IV}
Recently in \cite{Lashkari:2018oke}, the modular operator, as well as the associated flow, for a class of excited states has been obtained as a perturbation expansion. These states are cyclic and separating and they are created by invertible operators supported in a region of spacetime, acting on vacuum. In this section, our goal is to find the perturbative series expansion of modular zero modes for these class of excited states. 

By definition, we have:
\begin{align}
\Delta A_{0} \Delta^{-1} = A_{0}; \quad \implies [\Delta,A_{0}] = 0
\end{align}
Where $A_{0}$ is the zero mode and $\Delta$ is the modular operator in the excited state. We will now show, how to solve it perturbatively. To begin with, we look at the first order perturbation of $\Delta$ and $A_0$ around the vacuum state, $\Delta=\Delta^{(0)}+\Delta^{(1)}$ and similarly in the zero-mode $A_{0}=A^{(0)}_{0}+A^{(1)}_{0}$. 
\begin{align}\label{1st order commutation}
[\Delta^{(0)},A^{(1)}_{0}] = -[\Delta^{(1)},A^{(0)}_{0}]
\end{align}
The aim is now to solve equation \ref{1st order commutation} for $A^{1}_0$. We start with an ansatz, where the $A^1_0$ is expressed as a linear combination of all the non-zero eigenmodes of the $\Delta_0$\footnote{Also one could add linear combination of zero modes $\sum_{n}c^{(1)}_{n}(A^{(0)}_{0})^{(n)}$, of $\Delta_{0}$ to the ansatz. But we can always repackage such terms into redefining $A^{(0)}_{0}$  at each order.} and then solve for the coefficients of the eigenmodes. We will show how to do this explicitly. 
\begin{align}\label{ideal case}
A_{0}^{(1)} = \int^{\infty}_{-\infty}dk C_{k}B^{(0)}_{k} 
\end{align}

By definition the non zero modes $B^{(0)}_{k}$ satisfy the following equation, where $k$ is the eigen-value of the modular Hamiltonian $K$.
\begin{align}\label{transformation of non zero modes}
\Delta^{(0)}B^{(0)}_{k}\Delta^{(0)-1} = e^{k}B^{(0)}_{k}
\end{align}
Hence, it follows that the commutator $[\Delta^{(0)},B^{(0)}_{k}] = \left(e^{k}-1\right)B^{(0)}_{k}\Delta^{(0)}$. 
Using this, the (\ref{1st order commutation}) reduces to,
\begin{align}
\int^{\infty}_{-\infty}dk C_{k}\left(e^{k}-1\right)B^{(0)}_{k} = -[\Delta^{(1)},A^{(0)}_{0}]\Delta^{(0)-1}
\end{align}
If the non zero modes $B^{(0)}_{k}$, satisfy an orthogonality relation, then we could use it to explicitly obtain expressions for the $C_k$. 

As discussed in\cite{Faulkner:2017vdd} and reviewed in the introduction of this paper, we can form such orthogonal eigenmodes by taking Fourier transform of the modular transformed operators with support in a given region, as explicitly given in equation(\ref{mode defn}). The orthogonality condition satisfied by them are\footnote{See appendix (\ref{VIII}) for the proof.}:
\begin{align}\label{orthogonality}
\braket {\mathcal{O}^i_{k}\mathcal{O}^j_{k'}} = \frac{1}{2} \delta (k+k')\braket {\mathcal{O}^{i}_{k}\mathcal{O}^j} = \frac{1}{2} \delta (k+k')\braket {\mathcal{O}^i\mathcal{O}^{j}_{-k}}; \quad \mathcal{O}^i,\mathcal{O}^j \in \mathcal{A}
\end{align}
Using this, one can straightforwardly get the following expression for $C_{k}$, for any set of operators $\mathcal{O}_i,\mathcal{O}_j\in \mathcal{A}$
\begin{align}
C_{k} = \frac{\braket {[\Delta^{(1)},A^{(0)}_{0}]\Delta^{(0)-1}A_{k}^{(0)}}}{\braket {B^{(0)}_{k}\mathcal{O}^i}(1-e^{-k})}
\end{align}
Thus we end up with the following expression for $A_{0}^{(1)}$,
\begin{align}
A_{0}^{(1)} = \int^{\infty}_{-\infty}dk \frac{\braket {[\Delta^{(1)},A^{(0)}_{0}]\Delta^{(0)-1}A_{k}^{(0)}}}{\braket {B^{(0)}_{k}\mathcal{O}^i}(1-e^{-k})}B^{(0)}_{k} 
\end{align}
In a similar manner one can proceed to compute the next higher order corrections. For example, in second order, one could get the following equation,
\begin{align}
[\Delta^{(0)},A^{(2)}]=-[\Delta^{(2)},A^{(0)}_{0}]-[\Delta^{(1)},A^{(1)}_{0}]
\end{align}
Following the similar procedure used in first order correction, one could get the following
\begin{align}
A_{0}^{(2)} = \int^{\infty}_{-\infty}dk \frac{\braket {\left([\Delta^{(2)},A^{(0)}_{0}]+[\Delta^{(1)},A^{(1)}_{0}]\right)\Delta^{(0)-1}A_{k}^{(0)}}}{\braket {B^{(0)}_{k}\mathcal{O}^i}(1-e^{-k})}B^{(0)}_{k} 
\end{align}
One can easily generalize this to get the nth order expression for zero modes,
\begin{align}
A_{0}^{(n)} &= \int^{\infty}_{-\infty}dk \frac{\braket {\left([\Delta^{(n)},A^{(0)}_{0}]+[\Delta^{(n-1)},A^{(1)}_{0}]+\dots + [\Delta^{(1)},A^{(n-1)}_{0}]+[\Delta^{(0)},A^{(n)}_{0}] \right)\Delta^{(0)-1}A_{k}^{(0)}}}{\braket {B^{(0)}_{k}\mathcal{O}^i}(1-e^{-k})}B^{(0)}_{k} 
\end{align}
In this way, order by order we can perform the computation of zero modes in the class of excited state discussed in \cite{Lashkari:2018oke}. The expression for the perturbation of the modular operator $\Delta^{(n)}$, was obtained in \cite{Lashkari:2018oke}, and this method is reviewed in the appendix (\ref{IX}). 

Thus we can construct an infinite class of zero-modes (for each operator $\mathcal{O}^i$) of the excited state, using the information of the zero-modes of the vacuum modular Hamiltonian.
 
\section{Discussion}\label{V}

In this note, we have discussed some aspects of eigenmodes of the modular Hamiltonian, specifically in the context of AdS$_{3}$/CFT$_{2}$. We first showed in section (\ref{II}) that a set of OPE blocks, constructed from an OPE of two operators with non vanishing spin difference, form non-zero eigenmodes of the modular Hamiltonian for a single interval in the vacuum with eigenvalue being the spin difference, generalizing the scalar OPE blocks which form zero modes. In section (\ref{III}) we then gave a bulk description of these non zero mode in terms of bulk field integrated over a Lorentzian cylinder generated by the generators $K$ and $P_D$. 
In section (\ref{IV}) we provide a general setting to study zero modes in the class of excited state decribed in \cite{Lashkari:2018oke} around the vacuum in any dimension, as a perturbation around the zero-modes of the vacuum. \\

{\large\bf Comments on the self-adjointness of the modular operator}

Given a set of eigenmodes ($B_k$) of $\Delta_{\psi}$, we can construct eigenvectors from it, by acting it on the state $|\psi\rangle$. $|k;\psi\rangle \equiv B_{k}|\psi\rangle$.  
\begin{equation}
\Delta_{\psi}|k;\psi\rangle = e^{k}|k;\psi\rangle  
\end{equation}
Since $\Delta_{\psi}$ is a self-adjoint operator, it should have a complete set of orthogonal eigenvectors, with real eigenvalues. However the non-zero mode that we constructed out of the OPE block do not have real eigenvalues, since the eigenvalue of the modular Hamiltonian is pure imaginary ($il_{ij}$). We can also see that eigenvectors with different eigenvalues are not orthogonal, as shown in the appendix (\ref{X}). This means that these ope blocks do not lie in the domain over which the $\Delta$ is self-adjoint\footnote {For instance to take a quantum mechanical example of the particle in a box problem. We can always construct eigenfunctions $e^{ax}$, with $a$ complex. However, the Hamiltonian is self-adjoint over the domain of functions which vanish at the two ends of the box. Thus these eigenfunctions lie outside this domain. The eigenfunctions that do lie in this domain are the $sin(2\pi x/L)$ functions}. This also means that these OPE blocks cannot be used in the construction of the zero-modes for the excited states which we discussed in section(\ref{IV}). However, its interesting to note that a closely related object ($C_{k}$), can be constructed by analytic continuation in spins, which has real eigenvalues and are orthogonal, as is shown in the appendix (\ref{X}).\\

{\large\bf Comments on the dual of the excited state zero modes}

One potential application of our perturbative construction of the excited state zero-modes would be to use it to reconstruct a bulk scalar field in the dual theory perturbatively. The starting point of such a construction would be the work of \cite{Faulkner:2017vdd}, where the zero modes are identified with bulk operators localized on the RT surface in the following way,
\begin{align}
\mathcal{O}_{0}(x) = 4\pi \int_{\text{RT}}dY_{\text{RT}}\braket {\phi(Y_{\text{RT}})\mathcal{O}(x)} \phi(Y_{\text{RT}})
\end{align}

Formally, this relation may be inverted to reproduce local bulk field in terms of zero modes of that state. If we denote $f_{0}(X_{\text{RT}}|x)$ as the inverse of bulk-boundary correlator, then
\begin{align}
\phi(X_{\text{RT}}) = \int_{R} dx f_{0}(X_{\text{RT}}|x) \mathcal{O}_{0}(x)
\end{align}
Since, bulk modular operator commutes with the bulk field on the RT surface and bulk and boundary modular flows are same, it simply follows that,
\begin{align}
\Delta_{bulk}\phi(X_{\text{RT}})\Delta^{-1}_{bulk} = \Delta_{bdy}\phi(X_{\text{RT}}) \Delta^{-1}_{bdy} = \phi(X_{\text{RT}})
\end{align}
This further implies that,
\begin{align}
\phi(X_{\text{RT}})|_{\text{excited}} = \int_{R} dx f_{0}(X_{\text{RT}}|x)|_{\text{excited}} \Delta_{bdy,\text{excited}}\mathcal{O}_{0}(x)\Delta_{bdy,\text{excited}}^{-1}
\end{align}
In this note, we have shown how to construct zero modes in excited state perturbatively around the vacuum. Hence, $\Delta_{bdy,\text{excited}}\mathcal{O}_{0}(x)\Delta_{bdy,\text{excited}}^{-1} = \mathcal{O}_{0}^{(0)}(x) +\mathcal{O}_{0}^{(1)}(x)+ \dots$ Every terms can be computed explicitly as shown in this note, when vacuum modular operator and their modes are explicitly known. It would be interesting to see whether one could also obtain an expression for bulk field in a perturbation series around the vacuum AdS i.e $\phi(X_{\text{RT}})|_{\text{excited}} = \phi(X_{\text{RT}})|_{\text{vacuum}} + \phi(X_{\text{RT}})|_{\text{1st order}} + \dots$. But one should do it carefully. Since, the excited state is created by an excitation of operator on the vacuum, in the dual gravity side it changes the background geometry of vacuum AdS through the change of matter part in stress energy tensor. This change in geometry can be calculated explicitly by choosing a coordinate system like Fefferman-Graham coordinate where,
\begin{align}
ds^{2} = \frac{1}{z^{2}}\left[dz^{2}+g_{\mu\nu}dx^{\mu}dx^{\nu}\right]; \quad g_{\mu\nu} = g_{\mu\nu}^{(0),\text{vacuum AdS}}+z^{2}g_{\mu\nu}^{(2)}+z^{4}g_{\mu\nu}^{(4)}+\dots
\end{align}
Thus, in every order one could determine the location of RT surface from the corresponding back-reacted geometry\footnote{There are also recent works to probe the location of RT surface directly from the boundary \cite{Chen:2018rgz},\cite{Faulkner:2018faa}}. The knowledge of perturbative geometry at every order would help us to extract bulk-boundary propagator perturbatively by solving equation of motion for propagator and by taking one bulk field to boundary using AdS/CFT boundary condition. Thus at least formally one could get a perturbation of $f_{0}$ as $f_{0}(X_{\text{RT}}|x)|_{\text{excited}} = f_{0}(X_{\text{RT},0}|x)|_{\text{vac}}+f_{0}(X_{\text{RT},1}|x)|_{\text{1st order}}+ \dots$. And finally it generalizes the formal procedure of bulk reconstruction in any excited state on the RT surface perturbatively as $\phi(X_{\text{RT}})|_{\text{excited}} = \phi(X_{\text{RT},0})|_{\text{vacuum}} + \phi(X_{\text{RT},1})|_{\text{1st order}} + \dots$.
It would be nice, if this could be worked out explicitly atleast to first order. 

\vspace*{1ex}
\noindent{\bf Acknowledgment:}
SD would like to thank The Abdus Salam International Center for Theoretical Physics, Italy (Spring School on Super String Theory and Related Topics) for their warm hospitality during which, part of this work was completed. The work of SD was supported by a senior research fellowship(SRF) from CSIR.

 \appendix
\section{Proof of relation (\ref{main commutator})}\label{VI}
To obtain the desired commutator relation of modular Hamiltonian and OPE blocks, we use (\ref{ko commutator}) and get,
\begin{align}
&[K^{(R)},B^{ij}_{k}] \nonumber \\
&=  n_{ijk}\int_{y_{1}}^{y_{2}}d\zeta \int_{\bar{z}_{1}}^{\bar{z}_{2}}d\bar{\zeta}\left(\frac{(\zeta-y_{1})(y_{2}-\zeta)}{y_{2}-y_{1}}\right)^{h_{k}-1}\left(\frac{(\bar{\zeta}-\bar{y}_{1})(\bar{y}_{2}-\bar{\zeta})}{\bar{y}_{2}-\bar{y}_{1}}\right)^{\bar{h}_{k}-1} \left(\frac{(y_{2}-\zeta)(\bar{y}_{2}-\bar{\zeta})}{(\zeta-y_{1})(\bar{\zeta}-\bar{y}_{1})}\right)^{\frac{\Delta_{ij}}{2}} \nonumber \\
&\times\left(\frac{(y_{2}-\zeta)(\bar{\zeta}-\bar{y}_{1})}{(\zeta-y_{1})(\bar{y}_{2}-\bar{\zeta})}\right)^{\frac{l_{ij}}{2}} \frac{2\pi i}{y_{2}-y_{1}}[h(y_{2}+y_{1}-2\zeta)+(\zeta-y_{1})(y_{2}-\zeta)\partial_{\zeta}]\mathcal{O}_{k}(\zeta,\bar{\zeta}) \nonumber \\
& = n_{ijk}\int_{y_{1}}^{y_{2}}d\zeta \int_{\bar{z}_{1}}^{\bar{z}_{2}}d\bar{\zeta}\left(\frac{(\zeta-y_{1})(y_{2}-\zeta)}{y_{2}-y_{1}}\right)^{h_{k}-1}\left(\frac{(\bar{\zeta}-\bar{y}_{1})(\bar{y}_{2}-\bar{\zeta})}{\bar{y}_{2}-\bar{y}_{1}}\right)^{\bar{h}_{k}-1} \left(\frac{(y_{2}-\zeta)(\bar{y}_{2}-\bar{\zeta})}{(\zeta-y_{1})(\bar{\zeta}-\bar{y}_{1})}\right)^{\frac{\Delta_{ij}}{2}} \nonumber \\
&\times\left(\frac{(y_{2}-\zeta)(\bar{\zeta}-\bar{y}_{1})}{(\zeta-y_{1})(\bar{y}_{2}-\bar{\zeta})}\right)^{\frac{l_{ij}}{2}} \frac{2\pi i}{y_{2}-y_{1}}\times h(y_{2}+y_{1}-2\zeta)\mathcal{O}_{k}(\zeta,\bar{\zeta}) + \text{Total Derivative}(\int d\zeta \partial_{\zeta}(\dots)) \nonumber \\
& - n_{ijk}\int_{y_{1}}^{y_{2}}d\zeta \int_{\bar{y}_{1}}^{\bar{y}_{2}}d\bar{\zeta} (h_{k}-1) \left(\frac{(\zeta-y_{1})(y_{2}-\zeta)}{y_{2}-y_{1}}\right)^{h_{k}-2} \frac{(y_{2}+y_{1}-2\zeta)}{y_{2}-y_{1}} \left(\frac{(\bar{\zeta}-\bar{y}_{1})(\bar{y}_{2}-\bar{\zeta})}{\bar{y}_{2}-\bar{y}_{1}}\right)^{\bar{h}_{k}-1} \nonumber \\
&\times \left(\frac{(y_{2}-\zeta)(\bar{y}_{2}-\bar{\zeta})}{(\zeta-y_{1})(\bar{\zeta}-\bar{y}_{1})}\right)^{\frac{\Delta_{ij}}{2}} \left(\frac{(y_{2}-\zeta)(\bar{\zeta}-\bar{y}_{1})}{(\zeta-y_{1})(\bar{y}_{2}-\bar{\zeta})}\right)^{\frac{l_{ij}}{2}} \frac{2\pi i}{y_{2}-y_{1}}(\zeta-y_{1})(y_{2}-\zeta)\mathcal{O}_{k}(\zeta,\bar{\zeta}) \nonumber \\
& -n_{ijk}\int_{y_{1}}^{y_{2}}d\zeta \int_{\bar{z}_{1}}^{\bar{z}_{2}}d\bar{\zeta}\left(\frac{(\zeta-y_{1})(y_{2}-\zeta)}{y_{2}-y_{1}}\right)^{h_{k}-1}\left(\frac{(\bar{\zeta}-\bar{y}_{1})(\bar{y}_{2}-\bar{\zeta})}{\bar{y}_{2}-\bar{y}_{1}}\right)^{\bar{h}_{k}-1} \left(\frac{(y_{2}-\zeta)(\bar{y}_{2}-\bar{\zeta})}{(\zeta-y_{1})(\bar{\zeta}-\bar{y}_{1})}\right)^{\frac{\Delta_{ij}}{2}} \nonumber \\
&\times\left(\frac{(y_{2}-\zeta)(\bar{\zeta}-\bar{y}_{1})}{(\zeta-y_{1})(\bar{y}_{2}-\bar{\zeta})}\right)^{\frac{l_{ij}}{2}} \frac{2\pi i}{y_{2}-y_{1}} (y_{2}+y_{1}-2\zeta)\mathcal{O}_{k}(\zeta,\bar{\zeta}) \nonumber \\
& -n_{ijk}\int_{y_{1}}^{y_{2}}d\zeta \int_{\bar{z}_{1}}^{\bar{z}_{2}}d\bar{\zeta}\left(\frac{(\zeta-y_{1})(y_{2}-\zeta)}{y_{2}-y_{1}}\right)^{h_{k}-1}\left(\frac{(\bar{\zeta}-\bar{y}_{1})(\bar{y}_{2}-\bar{\zeta})}{\bar{y}_{2}-\bar{y}_{1}}\right)^{\bar{h}_{k}-1} \left(\frac{(y_{2}-\zeta)(\bar{\zeta}-\bar{y}_{1})}{(\zeta-y_{1})(\bar{y}_{2}-\bar{\zeta})}\right)^{\frac{l_{ij}}{2}} \nonumber \\
& \times \frac{\Delta_{ij}}{2} \left(\frac{(\bar{y}_{2}-\bar{\zeta})}{(\bar{\zeta}-\bar{y}_{1})}\right)^{\frac{\Delta_{ij}}{2}} \left(\frac{(y_{2}-\zeta)}{(\zeta-y_{1})}\right)^{\frac{\Delta_{ij}}{2}-1} \frac{y_{1}-y_{2}}{(\zeta-y_{1})^{2}}(\zeta-y_{1})(y_{2}-\zeta)\frac{2\pi i}{y_{2}-y_{1}}\mathcal{O}_{k}(\zeta,\bar{\zeta}) \nonumber \\
& -n_{ijk}\int_{y_{1}}^{y_{2}}d\zeta \int_{\bar{z}_{1}}^{\bar{z}_{2}}d\bar{\zeta}\left(\frac{(\zeta-y_{1})(y_{2}-\zeta)}{y_{2}-y_{1}}\right)^{h_{k}-1}\left(\frac{(\bar{\zeta}-\bar{y}_{1})(\bar{y}_{2}-\bar{\zeta})}{\bar{y}_{2}-\bar{y}_{1}}\right)^{\bar{h}_{k}-1}\left(\frac{(y_{2}-\zeta)(\bar{y}_{2}-\bar{\zeta})}{(\zeta-y_{1})(\bar{\zeta}-\bar{y}_{1})}\right)^{\frac{\Delta_{ij}}{2}}  \nonumber \\
& \times \frac{l_{ij}}{2} \left(\frac{(\bar{\zeta}-\bar{y}_{1})}{(\bar{y}_{2}-\bar{\zeta})}\right)^{\frac{l_{ij}}{2}} \left(\frac{(y_{2}-\zeta)}{(\zeta-y_{1})}\right)^{\frac{l_{ij}}{2}-1} \frac{y_{1}-y_{2}}{(\zeta-y_{1})^{2}}(\zeta-y_{1})(y_{2}-\zeta)\frac{2\pi i}{y_{2}-y_{1}}\mathcal{O}_{k}(\zeta,\bar{\zeta}) 
\end{align}
Here the total derivative term vanishes and other terms get cancelled except the terms involving $\Delta_{ij}$ and $l_{ij}$. Thus we have

\begin{align}
&[K^{(R)},B^{ij}_{k}] = \pi i (\Delta_{ij}+l_{ij})n_{ijk}\int_{y_{1}}^{y_{2}}d\zeta \int_{\bar{z}_{1}}^{\bar{z}_{2}}d\bar{\zeta}\left(\frac{(\zeta-y_{1})(y_{2}-\zeta)}{y_{2}-y_{1}}\right)^{h_{k}-1} \nonumber \\
&\times \left(\frac{(\bar{\zeta}-\bar{y}_{1})(\bar{y}_{2}-\bar{\zeta})}{\bar{y}_{2}-\bar{y}_{1}}\right)^{\bar{h}_{k}-1} \left(\frac{(y_{2}-\zeta)(\bar{y}_{2}-\bar{\zeta})}{(\zeta-y_{1})(\bar{\zeta}-\bar{y}_{1})}\right)^{\frac{\Delta_{ij}}{2}}\left(\frac{(y_{2}-\zeta)(\bar{\zeta}-\bar{y}_{1})}{(\zeta-y_{1})(\bar{y}_{2}-\bar{\zeta})}\right)^{\frac{l_{ij}}{2}}\mathcal{O}_{k}(\zeta,\bar{\zeta}) \\
& = \pi i (\Delta_{ij}+l_{ij})B^{ij}_{k}
\end{align}
 In a similar manner, one could also get the following,
\begin{align}
[K^{(R)},B^{ij}_{k}] = \pi i (l_{ij}-\Delta_{ij})B^{ij}_{k}
\end{align}

\section{Solution of curve equation (\ref{curve equation})}\label{VII}
In (\ref{curve equation}) we have the following set of equations that parametrize a curve ($x(s),z(s),t(s)$) generated by the flow of $H_{b,mod}$.
\begin{align}
&\frac{dt}{ds} \equiv \dot{t}=R^{2}-t^{2}-x^{2}-z^{2} \\
& \frac{dx}{ds} \equiv \dot{x}=-2xt \\
& \frac{dz}{ds} \equiv \dot{z}=-2zt
\end{align}
From the last two equations we have,
\begin{align}
\frac{dx}{dz} = \frac{x}{z}
\end{align}
This implies the $x-z$ plane of the curve as $x=az$ with some constant $a$. Along this curve, $t$ changes as $\dot{t} = R^{2}-t^{2}-\alpha^{2}z^{2}$, where $\alpha^{2}=1+a^{2}$. Now we can define the following quantities $u,v$ as,
\begin{align}
\dot{u} \equiv \dot{t} +\alpha\dot{z} = R^{2} - (t+\alpha z)^{2} = R^{2}-u^{2} \\
\dot{v} \equiv \dot{t} -\alpha\dot{z} = R^{2} - (t-\alpha z)^{2} = R^{2}-v^{2}
\end{align}
This implies,
\begin{align}
\frac{du}{R^{2}-u^{2}} = ds = \frac{dv}{R^{2}-v^{2}}
\end{align}
Solving this we get,
\begin{align}
\frac{u+R}{u-R} = Ae^{2Rs}; \quad A=\text{some constant} \\
\frac{v+R}{v-R} = Be^{2Rs}; \quad B=\text{some constant}
\end{align}
Let us define $\frac{A}{B} = \beta$ and we have,
\begin{align}
&(u+R)(v-R) = \beta (v+R)(u-R) \nonumber \\
& \implies uv + (v-u)R\frac{1+\beta}{1-\beta} - R^{2} = 0
\end{align}
Defining $\frac{1+\beta}{1-\beta} \equiv \gamma$ and putting $u=t+\alpha z$ and $v=t-\alpha z$ we will get,
\begin{align}
&t^{2} - \alpha^{2}z^{2} + 2\alpha zR\gamma - R^{2} = 0 \nonumber \\
&\implies t^{2} - \alpha^{2}\left(z-R\frac{\gamma}{\alpha}\right)^{2} = R^{2}(1-\gamma^{2})
\end{align}
Denoting $\eta = \frac{1}{\alpha}$, we get the desired curve equation of $z-t$ plane as in (\ref{z equation}), i.e
\begin{align}
\eta^{2}t^{2}-(z+R\gamma\eta)^{2} = \eta^{2}R^{2}(1-\gamma^{2})
\end{align}
In a similar way, one can find the corresponding curve equation in $x-t$ plane. In that case, we need to solve $\dot{t} = R^{2}-t^{2}-\tilde{\alpha}^{2}x^{2}$ and $\tilde{\alpha}\dot{x} = -2\tilde{\alpha} xt$, where $\tilde{\alpha}^{2} = 1+\frac{1}{a^{2}}$. Defining $\hat{\eta} = \frac{1}{\tilde{\alpha}}$ we will get the (\ref{x equation}), i.e
\begin{align}
\hat{\eta}^{2}t^{2}-(x+R\gamma\hat{\eta})^{2} = \hat{\eta}^{2}R^{2}(1-\gamma^{2})
\end{align}
It can be easily checked that $\eta^{2} +\hat{\eta}^{2} = 1$.
\section{Proof of orthogonality relation (\ref{orthogonality})}\label{VIII}
Using the mode definition (\ref{mode defn}), i.e
\begin{align}
B_{k}=\int_{-\infty}^{\infty}ds e^{-isk} e^{iKs}Be^{-iKs}; \quad B \in \mathcal{A}
\end{align}
We get the following
\begin{align}
\braket {B^{(0)}_{k}A^{(0)}_{k'}} &= \int_{-\infty}^{\infty}ds\int_{-\infty}^{\infty}ds' e^{-isk}e^{-is'k'} \braket {e^{iKs}Be^{-iK(s-s')}Ae^{-iKs'}}; \quad B,A \in \mathcal{A} \nonumber \\
& = \int_{-\infty}^{\infty}ds\int_{-\infty}^{\infty}ds' e^{-isk}e^{-is'k'} \braket {Be^{-iK(s-s')}A}
\end{align}
In the second step we use the purity condition of $\ket {\psi}$ i.e $K\ket {\psi} = 0$. Now using change of variables, $\tilde{s}=\frac{s+s'}{2}$ and $\tilde{s'} = \frac{s-s'}{2}$ we will get,
\begin{align}
\braket {B^{(0)}_{k}A^{(0)}_{k'}} &= \int_{-\infty}^{\infty}d\tilde{s}\int_{-\infty}^{\infty}d\tilde{s'} e^{-i\tilde{s}(k+k')}e^{-i\tilde{s'}(k-k')} \braket {Be^{-2iK\tilde{s'}}A} \nonumber \\
& =\frac{1}{2} \delta(k+k') \int_{-\infty}^{\infty}d\hat{s} e^{-i\hat{s}k} \braket {Be^{-iK\hat{s}}A} \nonumber \\
& = \frac{1}{2} \delta (k+k')\braket {B^{(0)}_{k}A} = \frac{1}{2} \delta (k+k')\braket {BA^{(0)}_{-k}}
\end{align}
In the second step we use the definition of delta function and also use $\hat{s} = 2\tilde{s'}$. In the final step we again use the definition of mode as in (\ref{mode defn}).
\section{Modular operator in excited states}\label{IX}
In \cite{Lashkari:2018oke}, it was shown that for any vacuum state(which is cyclic and separating) $\ket {\Omega}$, any excited state $\psi \ket {\Omega}$ and $\psi^{\dagger} \ket {\Omega}$, generated by invertible operators $\psi$ and $\psi^{\dagger}$, are cyclic and separating. For such states, one can define Tomita operator $S_{\psi}$ as,
\begin{align}
S_{\psi} = (\psi^{\dagger})^{-1} S_{\Omega} \psi
\end{align}
Where $S_{\Omega}$ is the Tomita operator in the vacuum. It can be easily checked that indeed $S_{\psi}$ and $S_{\psi}^{\dagger}$ satisfy 
\begin{align}
S_{\psi}A\psi\ket {\Omega} = A^{\dagger}\psi\ket {\Omega}; \quad \forall A \in \mathcal{A} \\
S_{\psi}^{\dagger}A'\psi\ket {\Omega} = A'^{\dagger}\psi\ket {\Omega}; \quad \forall A' \in \mathcal{A'}
\end{align}
Where $\mathcal{A}$ and $\mathcal{A'}$ are the algebra of bounded operators of a region and its compliment respectively. In this case, the modular operator $\Delta_{\psi}$ of the excited state $\psi \ket {\Omega}$ is,
\begin{align}
\Delta_{\psi} = S_{\psi}^{\dagger}S_{\psi} = \psi S_{\Omega}^{\dagger}\psi^{-1}(\psi^{-1})^{\dagger}S_{\Omega}\psi^{\dagger}
\end{align}
On the other hand one can also define a relative Tomita operator $S_{\phi\psi}$ between two state $\psi \ket {\Omega}$ and $\phi \ket {\Omega}$, then we have,
\begin{align}
S_{\phi\psi} = (\psi^{\dagger})^{-1}S_{\Omega}\phi^{\dagger}
\end{align}
Again it satisfies the definition of $S_{\phi\psi}$, i.e
\begin{align}
S_{\phi\psi}A\psi\ket {\Omega} = A^{\dagger}\phi\ket {\Omega}; \quad \forall A \in \mathcal{A} \\
S_{\phi\psi}^{\dagger}A'\psi\ket {\Omega} = A'^{\dagger}\phi\ket {\Omega}; \quad \forall A' \in \mathcal{A'}
\end{align}
Now, if we look at the relative Tomita operator between $\psi \ket {\Omega}$ and vacuum $\ket {\Omega}$, the Tomita operator $S_{\psi\Omega}$ and modular operator $\Delta_{\psi\Omega}$ become simplified as
\begin{align}\label{relative modular operator}
S_{\psi\Omega} = S_{\Omega}\psi^{\dagger}; \quad \Delta_{\psi\Omega} = S_{\psi\Omega}^{\dagger}S_{\psi\Omega} = \psi\Delta_{\Omega}\psi^{\dagger}
\end{align}
In \cite{Lashkari:2018oke}, the authors have noticed that instead of trying to calculate the flow with modular operator $\Delta_{\psi}$, it would be easy to do that with relative modular operator $\Delta_{\psi\Omega}$. Since the flows of modes under modular operator and relative modular operators are same, we can directly work with $\Delta_{\psi\Omega}A_{0}\Delta_{\psi\Omega}^{-1}$.
\begin{align}
\Delta_{\psi\Omega}^{is}A_{0}\Delta_{\psi\Omega}^{-is} = \Delta_{\psi}^{is}A_{0}\Delta_{\psi}^{-is}
\end{align}
 Hence, for perturbation of state $\ket {\psi}$ around the vacuum $\ket {\Omega}$, the operator $\psi(\mu)$ with perturbative parameter $\mu$ can be expanded around identity operator as
\begin{align}
\psi(\mu) = 1 + \mu\psi^{(1)} + \frac{\mu^{2}}{2}\psi^{(2)} + \dots
\end{align}
Putting this back in (\ref{relative modular operator}), we get the following,
\begin{align}
&\Delta^{(1)}_{\psi\Omega} = \psi^{(1)}\Delta_{\Omega} + \Delta_{\Omega}(\psi^{(1)})^{\dagger} \\
&\Delta^{(2)}_{\psi\Omega} = \psi^{(2)}\Delta_{\Omega} + \Delta_{\Omega}(\psi^{(2)})^{\dagger} + \psi^{(1)}\Delta_{\Omega}(\psi^{(1)})^{\dagger} \\
& \text{and so on} \dots \nonumber
\end{align}
Since we know $\Delta_{\Omega}$, we can compute all order computation and hence similarly we could get the zero modes of excited states at any order.

\section{Examples of CFT$_{2}$ non-zero modes and orthogonality}\label{X}
Let us consider the following non-local CFT$_{2}$ object by smearing a primary $\mathcal{O}_{k}$ of dimension $(h_{k},\bar{h}_{k})$:
\begin{align}
C_{k}(y_{1},\bar{y}_{1};y_{2},\bar{y}_{2}) =& n_{ijk}\int_{y_{1}}^{y_{2}}d\zeta \int_{\bar{z}_{1}}^{\bar{z}_{2}}d\bar{\zeta}\left(\frac{(\zeta-y_{1})(y_{2}-\zeta)}{y_{2}-y_{1}}\right)^{h_{k}-1}\left(\frac{(\bar{\zeta}-\bar{y}_{1})(\bar{y}_{2}-\bar{\zeta})}{\bar{y}_{2}-\bar{y}_{1}}\right)^{\bar{h}_{k}-1}\times \nonumber \\
&\left(\frac{(y_{2}-\zeta)(\bar{y}_{2}-\bar{\zeta})}{(\zeta-y_{1})(\bar{\zeta}-\bar{y}_{1})}\right)^{-i\alpha}\left(\frac{(y_{2}-\zeta)(\bar{\zeta}-\bar{y}_{1})}{(\zeta-y_{1})(\bar{y}_{2}-\bar{\zeta})}\right)^{-i\beta}\mathcal{O}_{k}(\zeta,\bar{\zeta}); \quad (\alpha,\beta \in \mathbb{R})
\end{align}
Using the similar steps as in section (\ref{II}), we would end up with,
\begin{align}
[K,C_{k}] = \pi \beta C_{k}
\end{align}
Therefore $C_{k}$s are a set of non zero modes of $K$ with real eigenvalues $\pi\beta$. From the definition of the modular modes (\ref{mode defn}) as a Fourier transform of the modular transformed operator $C(s)$, it follows that
\begin{align}
C(s) = \int^{\infty}_{-\infty}d\beta e^{is\beta}C_{k}
\end{align}
By putting the expression of $C_{k}$ we have
\begin{align}
C(s) =& n_{ijk}\int_{y_{1}}^{y_{2}}d\zeta \int_{\bar{y}_{1}}^{\bar{y}_{2}}d\bar{\zeta}\left(\frac{(\zeta-y_{1})(y_{2}-\zeta)}{y_{2}-y_{1}}\right)^{h_{k}-1}\left(\frac{(\bar{\zeta}-\bar{y}_{1})(\bar{y}_{2}-\bar{\zeta})}{\bar{y}_{2}-\bar{y}_{1}}\right)^{\bar{h}_{k}-1}\times \nonumber \\
&\left(\frac{(y_{2}-\zeta)(\bar{y}_{2}-\bar{\zeta})}{(\zeta-y_{1})(\bar{\zeta}-\bar{y}_{1})}\right)^{-i\alpha}\mathcal{O}_{k}(\zeta,\bar{\zeta}) \int^{\infty}_{-\infty}d\beta e^{i\beta\left[s-\ln\left(\frac{(y_{2}-\zeta)(\bar{\zeta}-\bar{y}_{1})}{(\zeta-y_{1})(\bar{y}_{2}-\bar{\zeta})}\right)\right]} 
\end{align}
The $\beta$ integral gives a Dirac delta function $\delta \left[s-\ln\left(\frac{(y_{2}-\zeta)(\bar{\zeta}-\bar{y}_{1})}{(\zeta-y_{1})(\bar{y}_{2}-\bar{\zeta})}\right)\right]$. It reduces to a single integral expression for $C(s)$.
\begin{align}
C(s) =& \int_{\bar{y}_{1}}^{\bar{y}_{2}}d\bar{\zeta}\left(\frac{(\bar{\zeta}-\bar{y}_{1})(\bar{y}_{2}-\bar{\zeta})(y_{2}-y_{1})}{[e^{-s/2}(\bar{\zeta}-\bar{y}_{1})+e^{s/2}(\bar{y}_{2}-\bar{\zeta})]^{2}}\right)^{\bar{h}_{k}-1} 
\left(\frac{(\bar{\zeta}-\bar{y}_{1})(\bar{y}_{2}-\bar{\zeta})}{\bar{y}_{2}-\bar{y}_{1}}\right)^{\bar{h}_{k}-1} \nonumber \\
&\left(e^{s/2}\frac{(\bar{y}_{2}-\bar{\zeta})}{(\bar{\zeta}-\bar{y}_{1})}\right)^{-2i\alpha}\mathcal{O}_{k}(\zeta,\bar{\zeta})|_{z=f(\bar{\zeta},s)}
\end{align}
Where $f(\bar{\zeta},s) = \frac{y_{2}(\bar{\zeta}-\bar{y}_{1})+e^{s}y_{1}(\bar{y}_{2}-\bar{\zeta})}{(\bar{\zeta}-\bar{y}_{1})+e^{s}(\bar{y}_{2}-\bar{\zeta})}$. Hence, by putting $s=0$, we can get the operator $C$
\begin{align}
C = & \int_{\bar{y}_{1}}^{\bar{y}_{2}}d\bar{\zeta}\left(\frac{(\bar{\zeta}-\bar{y}_{1})(\bar{y}_{2}-\bar{\zeta})(y_{2}-y_{1})}{[e^{-s/2}(\bar{y}_{2}-\bar{y}_{1}))^{2}}\right)^{\bar{h}_{k}-1} 
\left(\frac{(\bar{\zeta}-\bar{y}_{1})(\bar{y}_{2}-\bar{\zeta})}{\bar{y}_{2}-\bar{y}_{1}}\right)^{\bar{h}_{k}-1}\left(\frac{\bar{y}_{2}-\bar{\zeta}}{\bar{\zeta}-\bar{y}_{1}}\right)^{-2i\alpha}\mathcal{O}_{k}(f(\bar{\zeta},s=0),\bar{\zeta})
\end{align}
The existence of operator $C$ in CFT$_{2}$ suggests, the Fourier modes of its modular transform gives the non zero modes of modular Hamiltonian $K$ and hence $C_{k}$s satisfy the standard orthogonality condition (\ref{orthogonality}). A special case of $C_{k}$ is the identification of $\alpha$ and $\beta$ with $i\frac{\Delta_{ij}}{2}$ and $i\frac{l_{ij}}{2}$ respectively. It reduces to the OPE block $B^{ij}_{k}$, described in the section (\ref{II}) with non zero $l_{ij}$. It gives imaginary eigenvalue $2\pi il_{ij}$ as mentioned earlier. Thus, there exist two types of modular eigenmodes with real and imaginary eigenvalues. But since, $\Delta$ is a self-adjoint operator, only those set of eigenoperators with real eigenvalues form a complete set. 

Now we would like to see whether the OPE blocks $B^{ij}_{k}$ could be written in the following form
\begin{align}\label{wrong}
B^{ij}_{k}= \int ds e^{-isl_{ij}} B(s)
\end{align}
It yields,
\begin{align}
B(s) = \int^{\infty}_{-\infty}\rho (l_{ij})dl_{ij} e^{isl_{ij}}B^{ij}_{k}
\end{align}
Here $\rho (l_{ij})$ is spectral density function of $l_{ij}$, which is purely depend on the spectrum of the theory one consider. Even though one could ignore that $\rho$ formally(since it was just a inverse Fourier transform), the integral still diverges as one could check that. The divergence occurs due to the analytic continuation of real $\beta$ to imaginary $il_{ij}$ in this case. This implies, we cannot write $B^{ij}_{k}$ in this form of (\ref{wrong}). Nevertheless, it is well-known that the two point function involving $B^{ij}_{k}$ and $B^{i'j'}_{k'}$ gives precisely a conformal block $W_{h_{k}}^{ij,i'j'}$ in vacuum.
\begin{align}
&W_{h_{k}}^{ij,i'j'}(z,\bar{z})=\braket {B^{ij}_{k}(z_{1},z_{2}) B^{i'j'}_{k'}(z_{3},z_{4})} = k_{h}(z)k_{\bar{h}}(\bar{z})+k_{\bar{h}}(z)k_{h}(\bar{z}); \nonumber \\
&z=\frac{z_{12}z_{34}}{z_{13}z_{24}}; \quad k_{h}(z) = z^{h} {}_2F_1\left(h-h_{ij},h+h_{i'j'};2h;z\right); \quad h_{ij}=h_{i}-h_{j}, h_{i}=\frac{\Delta_{i}+l_{i}}{2}
\end{align}
Though such conformal block is orthogonal in $h_{k}$ and $h_{k'}$, it is not of the form $\propto \delta(l_{ij}+l_{i'j'})$. Hence, such OPE blocks are not desired modular modes to satisfy orthogonality in modular eigenvalues.

\end{document}